\documentclass[twocolumn,showpacs,aps,floatfix,superscriptaddress]{revtex4}
\pdfoutput=1 %This is for the ArXiv submission only
\usepackage{amsmath,amssymb,eucal,graphicx,float,epstopdf}

\begin{document}
\title{Mass Exchange Processes with Input}
\author{P.~L.~Krapivsky}
\affiliation{Department of Physics, Boston University, Boston, MA 02215, USA}

\begin{abstract}
We investigate a system of interacting clusters evolving through mass exchange and supplemented by input of small clusters. Three possibilities depending on the rate of exchange generically occur when input is homogeneous: continuous growth, gelation, and instantaneous gelation. We mostly study the growth regime using scaling methods. An exchange process with reaction rates equal to the product of reactant masses admits an exact solution which allows us to justify the validity of scaling approaches in this special case. We also investigate exchange processes with a localized input. We show that if the diffusion coefficients are mass-independent, the cluster mass distribution becomes stationary and develops an algebraic tail far away from the source.
\end{abstract}

\maketitle

\section{Introduction}
\label{sec:MEP}

Exchange processes underlie numerous phenomena. In natural sciences, exchange plays an important role in droplet growth via evaporation and re-condensation \cite{meakin}, island growth in deposition processes \cite{az}, and phase ordering \cite{ls,ajb,sm}. Exchange processes have been also applied to modeling segregation of heterogeneous populations \cite{ts}, studying the distribution of wealth \cite{IKR98}, mimicking growth of urban populations \cite{lr} and aggregation behaviors in job markets \cite{Sun14}.

The basic ingredient of exchange processes are clusters. We suppose that clusters are composed of a certain integer number of elemental building blocks (`monomers'); clusters are thus labelled by their masses, other characteristics (e.g., their shape) are not explicitly taken into account. We also assume that in each exchange event, a monomer is transferred from one cluster to another.  An exchange event can be separated into detachment of a monomer, subsequent transport to another cluster and attachment to it, so the comprehensive description of exchange processes can be complicated. We focus on the detachment-controlled limit when the time scale for transport and subsequent attachment is negligible. 

Denote by $A_j$ a cluster of mass $j$, that is, a cluster which is made of $j$ monomers. Symbolically, the mass exchange process can be represented as 
\begin{equation*}
A_i+A_j\mathop{\longrightarrow}^{K_{i,j}}A_{i\pm 1}+A_{j\mp 1}
\end{equation*}
We assume that a cluster is equally likely to gain or to lose mass in an interaction. This implies that the reaction rate matrix (`kernel' in short) is symmetric: $K_{i,j}=K_{j,i}$. A cluster disappears when its mass reaches zero. Thus the number of clusters decreases by 1 with probability 1/2 when strictly one of the reacting clusters is a monomer; if both reacting clusters are monomers, one of the monomers disappears and the other becomes a dimer. Exchange processes with rather general classes of homogeneous symmetric kernels, e.g., with kernels of the form $K_{i,j}=i^aj^b+i^bj^a$, have been studied in Refs.~\cite{KL02,EP03}. 

The density $c_m(t)$ of clusters containing $m$ monomers at time $t$ evolves according to the rate equation
\begin{equation}
\label{exchange:RE} 
\frac{dc_m}{dt} = \sum_{i,j} c_i c_j K_{i,j} \left[\delta_{m,i+1}+\delta_{m,i-1}-2\delta_{m,i}\right]
\end{equation}
This equation assumes perfect mixing, or equivalently, absence of spatial correlations. Furthermore, the total number of clusters is assumed to be infinite; otherwise, even if we ignore the locations of clusters, the system will eventually condense into a single cluster and close to the condensation time fluctuations in number of clusters become important, so the deterministic rate equation framework looses its validity. Even for infinite systems in homogeneous settings, the rate equation framework becomes inapplicable in low dimensions, particularly in one dimension (see \cite{ARS08,Evans14} and references therein). 

If the reaction kernel is the product of reactant masses, $K_{i,j}=ij$, the cluster densities $c_m(t)$ satisfy 
\begin{equation}
\label{exchange:prod}
\frac{dc_m}{d t}=M[(m+1)c_{m+1}-2mc_m+(m-1)c_{m-1}]
\end{equation}
and since the mass density $M=\sum_{j\geq 1} j c_j$  remains constant due to mass conservation, Eqs.~\eqref{exchange:prod} are linear. The linearity of rate equations distinguishes the product kernel from other kernels. If the system is initially composed of monomers, $c_m(t=0) = \delta_{m,1}$, the solution  reads
\begin{equation}
\label{prod:exact}
c_m(t) = \frac{t^{m-1}}{(1+t)^{m+1}}
\end{equation}
The corresponding total cluster density is  
\begin{equation}
\label{Nt:def}
N=\sum_{j\geq 1} c_j  = (1+t)^{-1}
\end{equation}
The cluster distribution \eqref{prod:exact} is the simplest exact result in the context of exchange processes \cite{IKR98,EP03}; it was originally discovered in earlier studies of birth-death processes (see \cite{Kendall49} and references therein). A few other exact solutions for exchange processes with more complicated kernels can be found in \cite{IKR98,EP03,book,FL03}.

Exchange processes help in classifying and interpreting emergent phenomena in various complex systems \cite{meakin,az,ls,ajb,sm,ts,IKR98,lr,Sun14}. Pure exchange process may provide an idealized description of binary interactions underlying the dynamics of such complex systems.  In addition, complex systems are usually open and driven. Perhaps the most ubiquitous example of an external drive is input; e.g., it is a key element in deposition processes and it naturally arises in phase ordering. The goal of this paper is to study the interplay between exchange and input. 

Pure exchange processes resemble aggregation processes \cite{similarity}. In the framework of aggregation, the influence of input has been studied. The outcome was somewhat surprising: For basic reaction kernels, aggregation processes with homogeneous input are {\em simpler} than corresponding pure aggregation processes, viz. in the most interesting large time limit the dominant part of the cluster mass distribution acquires a stationary form (see e.g. \cite{book,FL03,HH87,HT89,CRZ,Colm12}). There are exceptional kernels and whenever the stationary form is never reached, the emergent behaviors tend to be subtle (often they are characterized by more than one mass scale \cite{JM98,Colm12}). We shall show that in contrast to aggregation, exchange processes with spatially homogeneous input do not reach steady states. The emergent evolutionary behaviors, however, are usually qualitatively the same as for exchange processes without input \cite{EP03}. 

Our second goal is to study exchange processes with a localized source. A classification of possible behaviors in such processes appears unattainable---the behaviors depend on the reaction kernel, on the (mass-dependent) diffusion coefficients, and mathematically one has to study an infinite system of coupled non-linear partial differential equations. In the idealized situation when the diffusion coefficients are mass-independent, however, the models with a localized source are partly tractable. More precisely, the cluster mass distribution becomes stationary, at least in the most relevant three-dimensional setting, and this feature greatly simplifies analysis. 

Our analysis of mass exchange processes with input relies on scaling techniques. The emergence of scaling is difficult to prove, so it is useful to have exact results at least for simplest kernels which can be compared with outcomes of scaling analyses. It turned out that the model with product kernel remains solvable when we add homogeneous input. A few basic densities are still given by neat formulas, especially in the most natural setting when (i) only monomers are injected, and (ii) the system is initially empty. We will show that in this situation the total density reads 
\begin{equation}
\label{prod:total-input}
N  = \frac{1}{\sqrt{t^2+2}}\,\ln\frac{\sqrt{t^2+2} + t}{\sqrt{t^2+2} - t}
\end{equation}
while the density of monomers is
\begin{equation}
\label{prod:monomer-input}
c_1  = \frac{t}{t^2+2} +  \frac{1}{(t^2+2)^{3/2}}\,\ln\frac{\sqrt{t^2+2} + t}{\sqrt{t^2+2} - t}
\end{equation}

The remainder of this paper is organized as follows. In Sec.~\ref{sec:exact} we derive \eqref{prod:total-input}--\eqref{prod:monomer-input} and other exact results for the exchange process with product kernel and homogeneous input of monomers. This is a unique kernel for which an exact approach (based on the generating function technique) has been successful so far; another exactly soluble kernel (Sec.~\ref{sec:mon-dimer}) is degenerate as it only couples monomers and dimers. We are mostly interested in long time behaviors, however. Such behaviors can be explored using scaling techniques \cite{EP03,book,FL03}. Additionally, scaling techniques are applicable not just to the solvable product kernel, but to numerous kernels, even families of kernels. In Sec.~\ref{sec:scaling} we illustrate scaling techniques for exchange processes with three basic kernels: (i) product kernel $K_{i,j}=ij$,  (ii) constant kernel $K_{i,j}=1$, and (iii) sum kernel $K_{i,j}=i+j$. In Sec.~\ref{sec:GPK} we analyze exchange processes with a one-parameter family of kernels, namely the generalized product kernels $K_{i,j}=(ij)^{\lambda}$. We show that scaling methods are applicable to the generalized product kernels with homogeneity index $\lambda<\frac{3}{2}$. In the $\frac{3}{2}<\lambda\leq 2$, an infinite cluster (`gel') is formed at a finite time, while for $\lambda>2$ there is instantaneous gelation. In Sec.~\ref{sec:GPK} we also briefly discuss the extreme values of the homogeneity index: $\lambda=-\infty$ which is exactly soluble and $\lambda=2$ showing a multi-scaling behavior. In Sec.~\ref{sec:local} we study exchange processes driven by a localized source. We show that when the diffusion coefficients are mass-independent, the densities attain a steady state, at least in the most relevant three-dimensional setting, and we determine these densities for a few basic kernels. We also probe models with diffusion coefficients decaying as $(\text{mass})^{-1}$ and show that a stationary scaling mass distribution emerges when $\lambda<\frac{1}{2}$. We close (Sec.~\ref{sec:conl}) with a short discussion. 

\section{Exact Analysis}
\label{sec:exact}

Here we investigate the mass exchange process with product kernel and homogenous input using exact techniques. We shall assume that only monomers are injected. This is the most natural choice: The same elemental mass which is transferred in each exchange event is also added. The details of the input, however, do not affect qualitative behaviors (as long as clusters of only small masses are added). We shall also assume that the system is initially empty,
\begin{equation}
\label{IC}
c_m(t=0)=0,
\end{equation}
so that the mass density is equal to time,
\begin{equation}
\label{mass}
\sum_{j\geq 1} j c_j(t) = t,
\end{equation}
if only monomers are injected and if the monomer flux is set to unity.

In the case of the product kernel, $K_{i,j}=ij$, we thus arrive at the infinite set of rate equations
\begin{equation}
\label{prod:input}
\frac{dc_m}{d t} =  \delta_{m,1} + t[(m+1)c_{m+1}-2mc_m+(m-1)c_{m-1}]
\end{equation}

Equations \eqref{prod:input} are linear. In this respect, they are similar to Eqs.~\eqref{exchange:prod} describing the evolution without input. In contrast to Eqs.~\eqref{exchange:prod}, however, we now have an inhomogeneous system of equations with time-dependent coefficients. It is still possible to solve \eqref{prod:input} subject to the initial condition \eqref{IC}. Multiplying \eqref{prod:input} by $z^m$ and summing over all $m\geq 1$ we reduce an infinite set of rate equations \eqref{prod:input} to a single partial-differential equation 
\begin{equation}
\label{prod:GF}
\left[\frac{1}{t}\,\frac{\partial}{\partial t}-(1-z)^2\,\frac{\partial}{\partial z}\right]\mathcal{C} = \frac{z}{t}-c_1(t)
\end{equation}
for the generating function
\begin{equation}
\label{GF:def}
\mathcal{C}(z,t) = \sum_{m\geq 1} z^m c_m(t)
\end{equation}
Transforming variables $(z,t)$ to variables
\begin{equation}
\xi = \frac{t^2}{2} - \frac{1}{1-z}\,, \quad \eta =  \frac{t^2}{2} + \frac{1}{1-z}
\end{equation}
we recast \eqref{prod:GF} into
\begin{equation}
\label{prod:GF1}
2\,\frac{\partial \mathcal{C}}{\partial \xi} = \frac{1-\frac{2}{\eta - \xi}}{\sqrt{\xi + \eta}}-c_1\!\left(\!\sqrt{\xi + \eta}\right)
\end{equation}
Integrating \eqref{prod:GF1} yields
\begin{eqnarray}
\label{prod:GF-sol}
\mathcal{C}(\xi,\eta) &=& \int_{-\eta}^\xi \frac{du}{\sqrt{u+ \eta}}\,\left[\frac{1}{2} -\frac{1}{\eta - u}\right] \nonumber\\
 &-& \frac{1}{2}\int_{-\eta}^\xi du\,c_1\!\left(\!\sqrt{\xi + \eta}\right)
\end{eqnarray}
The choice of the lower integration limit in \eqref{prod:GF-sol} ensures $\mathcal{C}|_{\xi = -\eta} = 0$, which is equivalent to $\mathcal{C}(z,t=0) = 0$, i.e., to our choice of the initial condition, Eq.~\eqref{IC}.  

Computing the first integral in\eqref{prod:GF-sol} and returning to the original $(z,t)$ variables we obtain
\begin{eqnarray}
\label{prod:GF-final}
\mathcal{C}(z,t) &=& t - \int_0^t dv\,v c_1(v) \nonumber\\
 &-& \frac{1}{\sqrt{t^2+2\zeta}}\,\ln\frac{\sqrt{t^2+2\zeta} + t}{\sqrt{t^2+2\zeta} - t}
\end{eqnarray}
where $\zeta=(1-z)^{-1}$. 

\begin{figure}
\centering
\includegraphics[width=8cm]{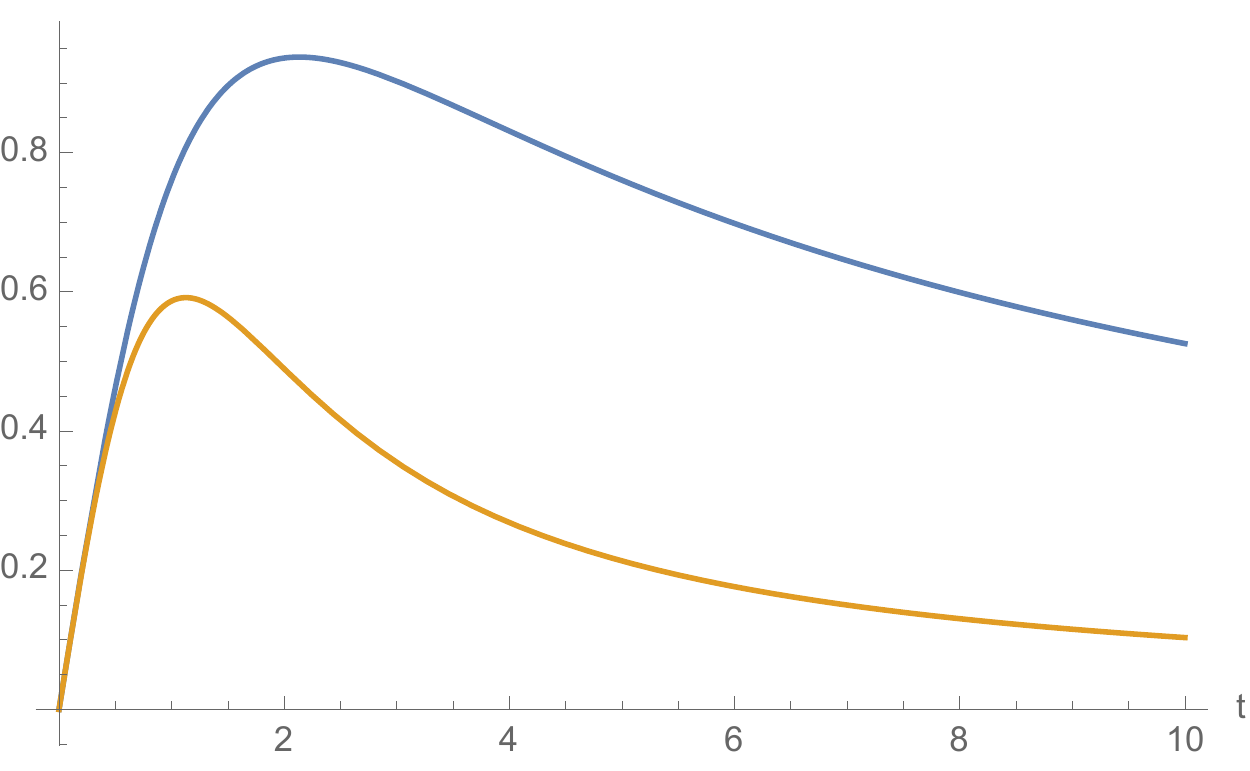}
\caption{Evolution of the total density $N(t)$ and the monomer density $c_1(t)$ in the mass exchange process with product kernel and homogeneous input. The top curve shows the total density  given by Eq.~\eqref{prod:total-input}; the bottom curve shows the monomer density given by Eq.~\eqref{prod:monomer-input}.}
\label{Fig:c1N}
\end{figure}

According to the definition \eqref{GF:def}, the generating function must vanish for $z=0$. Therefore 
\begin{equation}
\label{prod:mon}
\int_0^t dv\,v c_1(v) = t-\frac{1}{\sqrt{t^2+2}}\,\ln\frac{\sqrt{t^2+2} + t}{\sqrt{t^2+2} - t}
\end{equation}
leading to the announced expression \eqref{prod:monomer-input} for the monomer density (see also Fig.~\ref{Fig:c1N}). 

Using \eqref{prod:mon} we can re-write \eqref{prod:GF-final} as
\begin{eqnarray}
\label{prod:GF-final2}
\mathcal{C}(z,t) &=&  \frac{1}{\sqrt{t^2+2}}\,\ln\frac{\sqrt{t^2+2} + t}{\sqrt{t^2+2} - t} \nonumber\\
 &-& \frac{1}{\sqrt{t^2+2\zeta}}\,\ln\frac{\sqrt{t^2+2\zeta} + t}{\sqrt{t^2+2\zeta} - t}
\end{eqnarray}
Since $\mathcal{C}(z=1,t)=N(t)$  we obtain the total cluster density by specializing \eqref{prod:GF-final} to $z=1$. This gives \eqref{prod:total-input} shown on Fig.~\ref{Fig:c1N}. 

The generating function \eqref{prod:GF-final2} encapsulates all cluster densities. Expanding $\mathcal{C}(z,t)$ in power series in $z$ one can extract explicit expressions for cluster densities.  General expressions for cluster densities are similar to \eqref{prod:monomer-input}, viz.
\begin{equation}
\label{prod:cm-input}
c_m  = \frac{tP_m(t)}{(t^2+2)^m} +  \frac{Q_m(t)}{(t^2+2)^{m+1/2}}\,\ln\frac{\sqrt{t^2+2} + t}{\sqrt{t^2+2} - t}
\end{equation}
where $P_m(t)$ and $Q_m(t)$ are even polynomials of degree $2(m-1)$. For instance, 
\begin{equation*}
\begin{split}
& P_2 = \tfrac{1}{2}t^2 - \tfrac{1}{2}, \qquad \qquad \qquad \quad Q_2 = t^2 + \tfrac{1}{2}\\
& P_3 = \tfrac{1}{3}t^4 -  \tfrac{5}{6}t^2 - \tfrac{1}{2}, \qquad \qquad ~Q_3 = t^4 + t^2 + \tfrac{1}{2}\\
& P_4 = \tfrac{1}{4}t^6 - \tfrac{13}{12}t^4 -  \tfrac{31}{24}t^2 - \tfrac{5}{8}, \quad Q_4 = t^6+ \tfrac{3}{2}t^4 + \tfrac{3}{2}t^2 + \tfrac{5}{8}
\end{split}
\end{equation*}

\section{Scaling Analysis of Three Basic Kernels}
\label{sec:scaling}

In this section we employ scaling techniques. We start with the product kernel for which one can justify scaling using an exact solution from Sec.~\ref{sec:exact}. We then turn to two other simplest kernels, the constant kernel and the sum kernel, for which scaling appears to be the only available analytical technique. 

\subsection{Product Kernel}

Here we re-consider the product kernel, $K_{i,j}=ij$, but use scaling instead of exact methods. The typical mass is expected to grow indefinitely. For large masses, we can treat $m$ as a continuous variable and approximate the system of rate equations \eqref{prod:input} by the partial differential equation for the density $c=c_m(t)$:
\begin{equation}
\label{prod:cmt}
\frac{\partial c}{\partial t} = t\,\frac{\partial^2 (mc)}{\partial m^2}
\end{equation}
Balancing both sides of this equation yields $\frac{1}{t}\sim \frac{t}{m}$, so the typical mass grows according to $m\sim t^2$. 

To establish the asymptotic behavior of the density of monomers we sum all Eqs.~\eqref{prod:input} and obtain the exact rate equation for the total cluster density:
\begin{equation}
\label{prod:total}
\frac{dN}{d t} =  1 - tc_1
\end{equation}
This equation suggests that 
\begin{equation}
\label{prod:c1}
\lim_{t\to\infty} tc_1 = 1
\end{equation}
The rate equation for the density of monomers
\begin{equation}
\label{prod:c1t}
\frac{dc_1}{d t} =  1- 2t[c_1-c_2]
\end{equation}
suggests that $\lim_{t\to\infty} t(c_1-c_2) = 1/2$, which in conjunction with \eqref{prod:c1} give
\begin{equation}
\label{prod:c2}
\lim_{t\to\infty} tc_2 = \frac{1}{2}
\end{equation}
Similarly analyzing the rate equation for the density of dimers we get $\lim_{t\to\infty} tc_3 = \frac{1}{3}$, and generally for any {\em fixed} mass $k$ we have
\begin{equation}
\label{prod:ck}
\lim_{t\to\infty} tc_k = \frac{1}{k}
\end{equation}

Using \eqref{prod:ck} together with the growth law, $\langle m\rangle \sim t^2$, of the average mass we arrive at the following scaling form of the cluster distribution:
\begin{equation}
\label{cmt:scaling}
c_m(t) = \frac{1}{tm}\, F(\mu), \qquad \mu = \frac{m}{t^2}
\end{equation}
More precisely, the scaling form \eqref{cmt:scaling} is expected to hold in the scaling limit
\begin{equation}
\label{prod:scaling}
t\to\infty, \quad  m\to\infty, \quad \mu = \frac{m}{t^2} =\text{fixed}
\end{equation}
The small mass behavior \eqref{prod:ck} and the mass conservation law \eqref{mass} imply the following properties of the scaled mass distribution:
\begin{subequations}
\begin{align}
\label{FM:0}
F(0) &= 1\\
\label{FM:Int}
\int_0^\infty d\mu\,F(\mu) &= 1
\end{align}
\end{subequations}
The scaling form \eqref{cmt:scaling} together with \eqref{FM:0} lead to a rather unusual asymptotic behavior of the total cluster density:
\begin{equation}
\label{prod:N_asymp}
N\simeq \frac{2}{t}\,\ln t 
\end{equation}

\begin{figure}
\centering
\includegraphics[width=8cm]{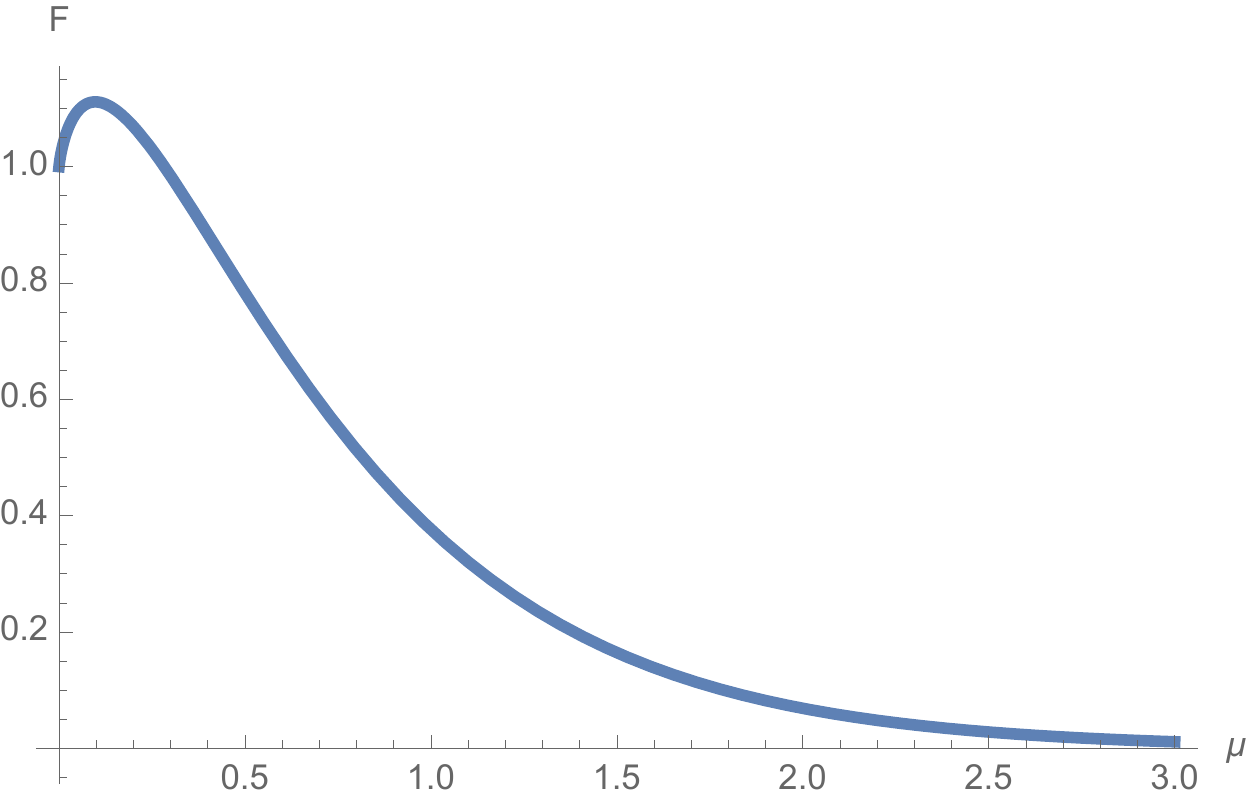}
\caption{The scaled mass distribution $F(\mu)$ given by Eq.~\eqref{F:sol_B} arising in the mass exchange process with product kernel and homogeneous input.}
\label{Fig:FM-prod}
\end{figure}

To determine the scaled mass distribution we plug the scaling form \eqref{cmt:scaling} into the governing partial differential equation \eqref{prod:cmt} and find that the scaled mass distribution satisfies an ordinary differential equation 
\begin{equation}
\label{FM:eq}
F''+2F'+\frac{1}{\mu}\,F = 0 
\end{equation}
where the differentiation with respect to $\mu$ is denoted by prime. Seeking solution in the form $F(\mu)=e^{-\mu} \Phi(\mu)$ we eliminate the term with first derivative:
\begin{equation}
\label{Phi:eq}
\Phi''+\left(\frac{1}{\mu}-1\right)\Phi = 0 
\end{equation}
WKB theory shows that there are two possible large $\mu$ asymptotic behaviors, $\mu^{-1/2}e^\mu$ and $\mu^{1/2}e^{-\mu}$. Only the latter asymptotic is acceptable. The boundary condition $F(0)=1$, equivalently $\Phi(0)=1$, then leads to the unique solution,  $\Phi=\mu[K_0(\mu)+K_1(\mu)]$, in terms of the modified Bessel functions $K_0$ and $K_1$. Thus
\begin{equation}
\label{F:sol_B}
F(\mu)=\mu\,e^{-\mu}\left[K_0(\mu)+K_1(\mu)\right]
\end{equation}
(see also Fig.~\ref{Fig:FM-prod}). Recalling that $K_1=-K_0'$, we can re-write \eqref{F:sol_B} as
\begin{equation}
\label{F:sol}
F= - \mu\,\frac{d}{d\mu}\left[e^{-\mu}K_0(\mu)\right]
\end{equation}
Plugging \eqref{F:sol} into \eqref{FM:Int} and performing the integration by parts we see that the validity of \eqref{FM:Int}  is equivalent to 
\begin{equation}
\label{K10}
\int_0^\infty d\mu\,e^{-\mu}K_0(\mu)=1
\end{equation}
This identity can be derived e.g. from the integral representation of the modified Bessel function $K_0$: 
\begin{equation}
\label{K0}
K_0(\mu) = \int_1^\infty \frac{du}{\sqrt{u^2-1}}\,e^{-\mu\,u}
\end{equation}

We also mention asymptotic behaviors of the scaled mass distribution. In the small mass limit ($\mu\to 0$) 
\begin{equation}
\label{FM:small}
F(\mu) -1 \simeq \mu\left(\ln\frac{2}{\mu}-1-\gamma\right)
\end{equation}
and in the large mass limit ($\mu\to\infty$)
\begin{equation}
\label{FM:large}
F(\mu) \simeq \sqrt{2\pi\mu}\,e^{-2\mu}
\end{equation}

For the product kernel we can use exact results to verify scaling predictions. The exact prediction \eqref{prod:monomer-input} for the monomer density gives
\begin{equation}
\label{prod:c1_asymp}
c_1 = \frac{1}{t} + \frac{1}{t^3}\,[2\ln t -2+ \ln 2] +\ldots
\end{equation}
when $t\gg 1$, so the scaling prediction \eqref{prod:c1} agrees with the exact prediction in the leading order. Similarly, the exact prediction \eqref{prod:total-input} or the total density,
\begin{equation}
\label{prod:N_asymptotic}
N\simeq \frac{1}{t}\,[2\ln  t + \ln 2] +  \frac{1}{t^3}\,[-\ln t + 1-\ln 2] +\ldots
\end{equation}
for $t\gg 1$, has the leading order term which agrees with the scaling prediction \eqref{prod:N_asymp}. 

To confirm the emergence of the mass scaling \eqref{prod:scaling} and deduce the scaled mass distribution it is convenient to re-write the exact prediction  \eqref{prod:GF-final2} for the generating function in the form
\begin{equation}
\label{GF_exact}
\sum_{m\geq 1} (1-z^m)c_m(t)= \frac{1}{\sqrt{t^2+2\zeta}}\,\ln\frac{\sqrt{t^2+2\zeta} + t}{\sqrt{t^2+2\zeta} - t}
\end{equation}
The scaling regime emerges when $\zeta \sim t^2$. Recalling that $\zeta=(1-z)^{-1}$, we write
\begin{equation}
\label{zs}
z = 1 - \frac{s}{t^2}
\end{equation}
and take the $t\to\infty$ limit while keeping $s$ finite. Combining \eqref{GF_exact} and \eqref{zs} we see the consistency of \eqref{cmt:scaling}--\eqref{prod:scaling}. Furthermore, plugging \eqref{zs} and \eqref{cmt:scaling} into \eqref{GF_exact} we obtain an integral equation 
\begin{equation}
\label{prod:IE}
\int_0^\infty \frac{d\mu}{\mu}\,(1-e^{-\mu s})F(\mu) = \sqrt{\frac{s}{s+2}}\,\ln\frac{\sqrt{s+2}+\sqrt{s}}{\sqrt{s+2}-\sqrt{s}}
\end{equation}
for the scaled mass distribution.  One can verify that the scaled mass distribution which we established using scaling methods is indeed the solution of the integral equation \eqref{prod:IE}. 

\subsection{Constant Kernel}
\label{sec:constant}

For the constant kernel $K_{i,j}=1$, the rate equations read
\begin{equation}
\label{const:input}
\frac{dc_m}{d t} =  \delta_{m,1} + N[c_{m+1}-2c_m+c_{m-1}]
\end{equation}
Summing \eqref{const:input} we obtain
\begin{equation}
\label{const:total}
\frac{dN}{d t} =  1 - Nc_1
\end{equation}

For large $t$ and $m$ we can approximate the system of rate equations \eqref{const:input} by the partial differential equation which happens to be the diffusion equation 
\begin{equation}
\label{const:cmt}
\frac{\partial c}{\partial \tau} = \frac{\partial^2 c}{\partial m^2}
\end{equation}
where we have use an auxiliary time variable:
\begin{equation}
\label{def:tau}
\tau = \int_0^t dt'\,N(t')
\end{equation}

The appropriate scaling form for the cluster mass distribution is
\begin{equation}
\label{cmt:scal}
c_m(\tau) = \tau^{-1/4}\, F(\mu), \qquad \mu = \frac{m}{\sqrt{\tau}}
\end{equation}
The form of the scaled mass, $\mu=m/\sqrt{\tau}$, is obvious from \eqref{const:cmt}. The time-dependent pre-factor in \eqref{cmt:scal} was chosen to ensure the validity of the asymptotic, $Nc_1\to 1$ when $\tau\to\infty$, which is implied by Eq.~\eqref{const:total}. Indeed, using \eqref{cmt:scal} we find that the cluster density $N(\tau)=\sum_{m\geq 1}c_m(\tau)$ grows according to
\begin{equation}
N = \tau^{1/4}\int_0^\infty d\mu\, F(\mu)
\end{equation}
Therefore $Nc_1$ is asymptotically constant and it equals to unity if  
\begin{equation}
\label{FF}
F(0)\int_0^\infty d\mu\, F(\mu) = 1
\end{equation}
Hence $N = [F(0)]^{-1}\tau^{1/4}$. Plugging this into \eqref{def:tau} we get
\begin{equation}
\label{t:tau}
t = \frac{4F(0)}{3}\,\tau^{3/4}
\end{equation}

Using \eqref{cmt:scal} we compute the mass density
\begin{equation}
\label{const:mass}
\sum_{j\geq 1} j c_j(t) = \tau^{3/4}\int_0^\infty d\mu\, \mu F(\mu) = t
\end{equation}
Comparing \eqref{t:tau} and \eqref{const:mass} we obtain
\begin{equation}
\label{FFF}
\frac{4F(0)}{3} = \int_0^\infty d\mu\, \mu F(\mu)
\end{equation}

To complete the analysis we must find the scaled mass distribution $F(\mu)$. Plugging \eqref{cmt:scal} into \eqref{const:cmt} leads to an ordinary differential equation 
\begin{equation}
\label{FM2}
F''+\tfrac{1}{2}\mu F'+\tfrac{1}{4}\,F = 0 
\end{equation}
Demanding that $F(\mu)$ decays sufficiently rapidly we arrive at a one-parameter family of solutions
\begin{equation}
\label{FM:1d}
F = C\,\sqrt{\mu}\,\exp(-\mu^2/8)\, K_{1/4}(\mu^2/8)
\end{equation}
involving the modified Bessel function $K_{1/4}$. 

To fix the amplitude $C$ we use relations \eqref{FF} and \eqref{FFF}. In computations, an integral representation of the modified Bessel function $K_{1/4}$ proves useful. This integral representation is obtained by specializing 
\begin{equation*}
K_\nu(z) = \frac{\sqrt{\pi}}{\Gamma\big(\nu+\frac{1}{2}\big)}\left(\frac{z}{2}\right)^\nu \int_1^\infty \frac{du}{\sqrt{u^2-1}}\,(u^2-1)^\nu \, e^{-z\,u}
\end{equation*}
to $\nu=1/4$. Using above integral representation one can compute the two-parameter family of integrals, 
\begin{equation}
\label{Kbn:def}
K(\beta,\nu):=\int_0^\infty dz\,z^{\beta-1}\,e^{-z}\,K_\nu(z),
\end{equation}
with arbitrary $\beta, \nu$ satisfying $\beta>\nu$. One finds
\begin{eqnarray}
\label{Kbn}
K(\beta,\nu) &=& \frac{\sqrt{\pi}\,\Gamma(\beta+\nu)}{2^\nu\,\Gamma\big(\nu+\frac{1}{2}\big)}
\int_1^\infty du\,\frac{(u-1)^{\nu-\frac{1}{2}}}{(u+1)^{\beta+\frac{1}{2}}}\nonumber\\
&=& \frac{\sqrt{\pi}\,\Gamma(\beta+\nu)\,\Gamma(\beta-\nu)}{2^\beta\,\Gamma\big(\beta+\frac{1}{2}\big)}
\end{eqnarray}
As a check we note that $K(1,0)$ reduces to the integral in \eqref{K10} and the right-hand side of \eqref{Kbn} is equal to unity for $(\beta,\nu) = (1,0)$. 

Using \eqref{FM:1d} and writing $\mu=\sqrt{8z}$ we express the integrals appearing in \eqref{FF} and \eqref{FFF} through the integrals $K(\beta,\nu)$, namely
\begin{equation}
\label{FK}
\begin{split}
\int_0^\infty d\mu\, F(\mu) &= C\cdot 2^{5/4} K\big(\tfrac{3}{4}, \tfrac{1}{4}\big)\\
\int_0^\infty d\mu\, \mu F(\mu) &= C\cdot 2^{11/4} K\big(\tfrac{5}{4}, \tfrac{1}{4}\big)
\end{split}
\end{equation}
Combining \eqref{FK} with \eqref{Kbn} and adding $F(0)$ we get
\begin{subequations}
\begin{align}
\label{F0}
F(0) &= C\,\frac{\pi\sqrt{2}}{\Gamma\big(\frac{3}{4}\big)}\\
\label{F1}
\int_0^\infty d\mu\, F(\mu) &= C\,\frac{\pi\sqrt{2}}{\Gamma\big(\frac{5}{4}\big)}\\
\label{F2}
\int_0^\infty d\mu\, \mu F(\mu) &= C\,\frac{\pi\sqrt{2}}{\Gamma\big(\frac{3}{4}\big)}\times \frac{4}{3}
\end{align}
\end{subequations}
Equations \eqref{F0} and \eqref{F2} show that \eqref{FFF} is manifestly satisfied. Using \eqref{F0} and \eqref{F1} we see that to ensure the validity of \eqref{FF} we must choose
\begin{equation}
\label{C:1d}
C= \frac{1}{2^{5/4}\,\sqrt{\pi}}= 0.237212499\ldots
\end{equation}

The asymptotic behaviors of the scaled mass distribution are
\begin{equation}
\label{small}
F(\mu)\simeq F(0)=\frac{\sqrt{\pi}}{2^{3/4}\,\Gamma\big(\frac{3}{4}\big)}=0.860039988\ldots
\end{equation}
when $\mu\to 0$ and 
\begin{equation}
\label{large}
F(\mu) \simeq \frac{1}{2^{1/4}\sqrt{\mu}}\,e^{-\mu^2/4}
\end{equation}
when $\mu\to\infty$. The latter formula implies that
\begin{equation}
c_m(\tau)\simeq \frac{1}{2^{1/4}\sqrt{m}}\,\exp\!\left[-\frac{m^2}{4\tau}\right]
\end{equation}
for $m\gg \sqrt{\tau}$. 

Finally, we give asymptotic formulas for the total density and the monomer density through the original time variable:
\begin{equation}
\label{sum:1/3}
N=At^{1/3}, \qquad c_1=A^{-1}t^{-1/3}
\end{equation}
where the amplitude is given by
\begin{equation*}
A=\frac{6^{1/3} \left[\Gamma\big(\frac{3}{4}\big)\right]^{4/3}}{\pi^{2/3}} = 1.1108674179\ldots
\end{equation*}

\subsection{Sum Kernel}

For the sum kernel $K_{i,j}=i+j$, the rate equations read 
\begin{eqnarray}
\label{prod:sum}
\frac{dc_m}{d t} &=& N[(m+1)c_{m+1}-2mc_m+(m-1)c_{m-1}]\nonumber\\
&+& t[c_{m+1}-2c_m+c_{m-1}] +  \delta_{m,1} 
\end{eqnarray}

The total cluster density satisfies
\begin{equation}
\frac{d N}{d t} = 1-tc_1
\end{equation}
leading to the same asymptotic relation \eqref{prod:c1} for the monomer density as in the case of the product kernel. 

In the scaling regime, the cluster density $c=c_m(t)$ obeys the partial differential equation
\begin{equation}
\label{sum:cmt}
\frac{\partial c}{\partial t} = N\,\frac{\partial^2 (mc)}{\partial m^2} + t\,\frac{\partial^2 c}{\partial m^2} 
\end{equation}
Balancing the left-hand side with the first term on the right-hand side gives $m\sim Nt$, while balancing with the second term leads to $m\sim t$. These estimates for the typical mass are comparable, and thence both terms on the right-hand side remain asymptotically relevant, if $N$ saturates at a certain finite value $N_\infty$. Below we compute this value and find
\begin{equation}
\label{N:sum_kernel}
\lim_{t\to\infty} N(t) = N_\infty = 1
\end{equation}
but now we merely suppose that the cluster density reaches a finite value $N_\infty$. Since $m\sim t$, the cluster mass distribution is sought in the scaling form
\begin{equation}
\label{sum:scaling}
c_m(t) = \frac{1}{t}\, F(\mu), \qquad \mu = \frac{m}{t}
\end{equation}
Combining \eqref{sum:cmt} and \eqref{sum:scaling} we arrive at 
\begin{equation*}
F+\mu F' +[F+N_\infty \mu F]'' = 0
\end{equation*}
Integrating once we obtain $\mu F +[F+N_\infty \mu F]' = 0$, or equivalently 
\begin{equation*}
[1+N_\infty \mu] F' +[N_\infty + \mu] F= 0
\end{equation*}
Integrating this equation gives a family of solutions
\begin{equation}
\label{FNC}
F = \frac{C}{(1+N_\infty \mu)^{1-1/N_\infty^2}}\,e^{-\mu/N_\infty}
\end{equation}
with two un-determined constants: $C$ and $N_\infty$. 

We should impose three conditions on the scaled mass distribution:
\begin{subequations}
\begin{align}
\label{F0:sum}
F(0) &= 1\\
\label{F1:sum}
\int_0^\infty d\mu\, F(\mu) &= N_\infty\\
\label{F2:sum}
\int_0^\infty d\mu\, \mu F(\mu) &= 1
\end{align}
\end{subequations}
Equation \eqref{F0:sum} ensures the validity of \eqref{prod:c1}; Eq.~\eqref{F1:sum} follows from $\lim_{t\to\infty} \sum_{m\geq 1}c_m(t) = N_\infty$; the last condition, Eq.~\eqref{F2:sum}, is the conservation law $\sum_{m\geq 1}c_m(t) = t$. Equation \eqref{F0:sum} obviously leads to $C=1$. Further, one can verify that the only choice of $N_\infty$ which agrees with Eqs.~\eqref{F1:sum}--\eqref{F2:sum} is $N_\infty=1$ as it was stated in \eqref{N:sum_kernel}. Thus \eqref{FNC} simplifies to
\begin{equation}
\label{F:sum}
F = e^{-\mu}
\end{equation}

\section{Generalized Product Kernels}
\label{sec:GPK}

In this section we consider exchange processes with generalized product kernels, $K_{i,j}=(ij)^\lambda$, and homogeneous input of monomers. The rate equations read
\begin{equation}
\label{GP:input}
\frac{dc_m}{d t} =  M_\lambda [(m+1)^\lambda c_{m+1} -2m^\lambda c_m +(m-1)^\lambda c_{m-1}]
\end{equation}
for $m>1$, while the monomer density satisfies
\begin{equation}
\label{GP:monomer}
\frac{dc_1}{d t} = 1 +  M_\lambda [2^\lambda c_2 -2 c_1]
\end{equation}
Here $M_\lambda$ is the $\lambda^\text{th}$ moment of the cluster mass distribution:
\begin{equation}
\label{GP:moment}
M_\lambda(t) = \sum_{m\geq 1} m^\lambda c_m(t)
\end{equation}
Note that the cluster density is the zeroth moment and the mass density is first moment of the cluster mass distribution: $N=M_0$ and $M=M_1$. 

For large $t$ and $m$ we replace the system of rate equations \eqref{GP:input} by the partial differential equation \begin{equation}
\label{GP:cmt}
\frac{\partial c}{\partial \tau} = \frac{\partial^2 (m^\lambda c)}{\partial m^2}
\end{equation}
where $\tau$ is an auxiliary time variable 
\begin{equation}
\label{GP:tau}
\tau = \int_0^t dt'\,M_\lambda(t')
\end{equation}

We seek the cluster mass distribution in the scaling form
\begin{equation}
\label{GP:scal}
c_m(\tau) = \tau^{-b}\, F(\mu), \qquad \mu = \frac{m}{\tau^a}
\end{equation}
The form of the governing equation \eqref{GP:cmt} fixes the exponent $a$ which determines the growth of the typical mass:
\begin{equation}
\label{GP:a}
a = \frac{1}{2-\lambda}
\end{equation}
To find another exponent let us establish the growth law of $M_\lambda$. Plugging \eqref{GP:scal} into \eqref{GP:moment} we get $M_\lambda \sim \tau^{(1+\lambda)a-b}$. Combining this result with \eqref{GP:tau} yields
\begin{equation}
\label{GP:tt1}
t\sim \tau^{1+b-(1+\lambda)a}
\end{equation}
Computing the mass density gives another relation between $t$ and the modified time $\tau$:
\begin{equation}
\label{GP:tt2}
t=M(t)\sim \tau^{2a-b}
\end{equation}
Relations \eqref{GP:tt1} and \eqref{GP:tt2} are compatible only when
\begin{equation}
\label{GP:b}
b = \frac{1+2\lambda}{2(2-\lambda)}
\end{equation}

The expressions \eqref{GP:a} and \eqref{GP:b} for the scaling exponents already tell us that the above analysis is inapplicable when $\lambda\geq 2$. For $\lambda>2$ the mathematical framework becomes ill-defined due to instantaneous gelation \cite{EP03}. 

Instantaneous gelation is a rather counter-intuitive phenomenon implying that an infinite cluster (`gel') nucleates at time  $t=+0$. Instantaneous gelation was originally discovered in the context of aggregation \cite{dom,hez,spouge,van87}, see \cite{Colm:IG,Leyvraz12} for recent work. Instantaneous gelation also arises in addition processes \cite{BK91} where it was rigorously proved \cite{Non99}. Instantaneous gelation was mostly studied in the mean-field framework, but it may also occur in one dimension  \cite{Evans14,Evans12} where spatial effects are crucial and the mean-field rate equation approach is inapplicable.

In the range $\frac{3}{2}\leq \lambda< 2$, one can employ scaling analysis using the auxiliary time and considering the usual $\tau\to\infty$ regime. One cannot use the original time variable, however. This can be appreciated by using \eqref{GP:a}, \eqref{GP:b} to re-write \eqref{GP:tt2} as $t\sim \tau^{(3-2\lambda)/(4-2\lambda)}$ indicating that consistent behaviors occur only when $\lambda<\frac{3}{2}$. In the $\frac{3}{2}<\lambda< 2$ range the system actually exhibits gelation, namely the gel forms in a finite time. Interestingly, gelation is complete meaning that the entire mass is suddenly transformed into gel. The analysis of the behavior in the gel regime is essentially the same as in un-driven exchange processes with the same kernels \cite{EP03}. Indeed, the most interesting behavior arises close to the gelation point, and the mass density is finite as in the un-driven case. Therefore let us focus on the $\lambda<\frac{3}{2}$ range where clusters grow indefinitely and the emergent scaling behavior differs from the un-driven case. 

Instead of the scaling form \eqref{GP:scal}, let us use an equivalent, but slightly more convenient, scaling form
\begin{equation}
\label{GP:scaling}
c_m(\tau) = \tau^{-a/2}\, m^{-\lambda}\,\Phi(\mu), \quad \mu = \frac{m}{\tau^a}\,, \quad  a = \frac{1}{2-\lambda}
\end{equation}
Plugging this scaling form into the governing equation \eqref{GP:cmt} we arrive at the ordinary differential equation
\begin{equation}
\label{GP:FM}
\mu^\lambda \Phi''+a\left[\mu \Phi'+\tfrac{1}{2}\Phi\right] = 0 
\end{equation}
Demanding that $\Phi(\mu)$ decays sufficiently rapidly we arrive at a one-parameter family of solutions
\begin{equation}
\label{GP:FM_sol}
\Phi(\mu) = C\, _{1}F_1\big(a/2, 1-a; - a^2 \mu^{1/a}\big)
\end{equation}
in terms of the (confluent) hypergeometric function
\begin{equation*}
_{1}F_1(\alpha, \beta; z) = \frac{\Gamma(\beta)}{\Gamma(\alpha)}\,
\sum_{n\geq 0}\frac{\Gamma(\alpha + n)}{\Gamma(\beta + n)}\,\frac{z^n}{n!}
\end{equation*}

Using \eqref{GP:scaling} and \eqref{GP:FM_sol} we compute the $\lambda^\text{th}$ moment:
\begin{equation}
\label{GP:ML}
M_\lambda = C\,\tau^{a/2} I_1
\end{equation}
with
\begin{equation}
\label{I1}
I_1 = a\int_0^\infty dz\,z^{a-1}\,_{1}F_1(a/2, 1-a; - a^2 z)
\end{equation}
Similarly we compute the first moment, the mass density:
\begin{equation}
\label{GP:M1}
M = t = C\,\tau^{1-a/2} I_2
\end{equation}
with
\begin{equation}
\label{I2}
I_2 = a\int_0^\infty dz\,_{1}F_1(a/2, 1-a; - a^2 z)
\end{equation}

Equation \eqref{GP:tau} tells us that $\frac{dt}{d\tau}=1/M_\lambda$ which is combined with \eqref{GP:ML} to give
\begin{equation*}
\frac{dt}{d\tau}= \tau^{-a/2}\, \frac{1}{CI_1}
\end{equation*}
Differentiating \eqref{GP:M1} with respect to $\tau$ we obtain another expression
\begin{equation*}
\frac{dt}{d\tau}= \tau^{-a/2}\, (1-a/2) CI_2
\end{equation*}
for the same derivative. These two formulas are compatible when 
\begin{equation}
C = \frac{1}{\sqrt{(1-a/2) I_1 I_2}}
\end{equation}

Note also the behavior of the total cluster density 
\begin{equation}
\label{GP:M0}
N = CI_0\,\tau^{(1-2\lambda)/(4-2\lambda)}=CI_0\left(\frac{t}{CI_2}\right)^{\frac{1-2\lambda}{3-2\lambda}}
\end{equation}
with
\begin{equation}
\label{I0}
I_0 = a\int_0^\infty \frac{dz}{z^a}\,\,_{1}F_1(a/2, 1-a; - a^2 z)
\end{equation}

Let us look more carefully at a few particular values of the homogeneity index.

\subsection{$\lambda=\frac{1}{2}$}
\label{sec:1/2}

The scaled mass distribution substantially simplifies when the homogeneity index is $\lambda=\frac{1}{2}$. In this case the parameters of the confluent hypergeometric function coincide ($\frac{a}{2}=1-a=\frac{1}{3}$), so the function 
becomes exponential. More precisely, 
\begin{equation}
\label{GP:2d}
\Phi(\mu) = C\,\exp\!\left[-\left(\tfrac{2}{3}\right)^2 \mu^{3/2}\right]
\end{equation}
with
\begin{equation}
\label{C:2d}
C = \left(\frac{2}{3}\right)^{1/6} \left[\Gamma\left(\frac{2}{3}\right)\right]^{-1/2}=0.8031990959\ldots
\end{equation}
The total cluster density in this case saturates at
\begin{equation}
\label{N:2d}
N_\infty = \frac{\pi}{3}\,\left[\frac{2}{\Gamma\big(\frac{2}{3}\big)}\right]^{3/2}= 1.87970144\ldots
\end{equation}

\subsection{Exchange process in $d$ dimensions}

Consider an exchange process supplemented by input of monomers in $d$ spatial dimensions. We assume that clusters are spherical (say because of the surface tension) and that in an exchange event a monomer on the surface of one cluster migrates to the surface of another cluster. This is similar to the model proposed by Schelling \cite{ts} to account for segregation. Another interpretation is an Ising model with infinite-range zero-temperature Kawasaki spin-exchange dynamics \cite{EP03}. The precise relation between all these models is rather subtle, e.g., there is no clear surface tension in the simplest versions of the Schelling and Ising-Kawasaki models, the shape of clusters apparently reflects the lattice structure \cite{EP03}. Coming back to our exchange process we note that the corresponding kernel is the product of the surface areas of the clusters participating in exchange: $K_{i,j}=\sigma_i\sigma_j$. Expressing the area of the cluster and its mass through the radius, $\sigma\sim R^{d-1}$ and $m\sim R^d$, we conclude that $\sigma_m \sim m^{1-1/d}$, so this exchange process is described by the generalized product kernel with $\lambda=1-\frac{1}{d}$. 

In one dimension $\lambda=0$, so the exchange process has constant kernel (Sec.~\ref{sec:constant}). In particular, the scaled mass distribution is given by \eqref{FM:1d} and \eqref{C:1d}, the total cluster density grows as $t^{1/3}$, Eq.~\eqref{sum:1/3}. In two dimensions $\lambda=\frac{1}{2}$, so we recover the exchange process described in the previous subsection \ref{sec:1/2}. The scaled mass distribution in two dimensions is given by \eqref{GP:2d}--\eqref{C:2d} and the total cluster density saturates at a constant value, Eq.~\eqref{N:2d}. In three dimensions $\lambda=\frac{2}{3}$, the scaled mass density becomes
\begin{equation}
\label{GP:3d}
\Phi(\mu) = C\, _{1}F_1\big(\tfrac{3}{8}, \tfrac{1}{4}; - \tfrac{9}{16} \mu^{4/3}\big)
\end{equation}
and the total cluster density decays as $t^{-1/5}$. Generally in $d$ dimensions: $N\sim t^{(2-d)/(2+d)}$.

\subsection{$\lambda=-\infty$}
\label{sec:mon-dimer}

For this smallest value of the homogeneity index, the scaling approach is no longer applicable. Yet the analysis of this case is simple since (i) only monomers participate in exchange, and (ii) in addition to monomers injected into the system, only dimers can be formed. 

The rate equations \eqref{GP:input} give a single non-trivial equation
\begin{equation}
\label{extreme:c2}
\frac{dc_2}{dt} = c_1^2
\end{equation}
while \eqref{GP:monomer} reduces to 
\begin{equation}
\label{extreme:c1}
\frac{dc_1}{d t} = 1 - 2 c_1^2
\end{equation}
Solving \eqref{extreme:c2}--\eqref{extreme:c1} subject to $c_1(0)=c_2(0)=0$ we obtain
\begin{equation}
\label{extreme:c12}
c_1 = \frac{1}{\sqrt{2}}\,\tanh T, \quad c_2 = \frac{1}{2\sqrt{2}}\,[T-\tanh T]
\end{equation}
with $T=t\sqrt{2}$. 

\subsection{$\lambda=2$}

The value $\lambda=2$ of the homogeneity index is another extreme: It separates the $\frac{3}{2}<\lambda\leq 2$ region where the gel is formed from the $\lambda>2$ region where instantaneous gelation occurs. The simplest way to convince oneself in the emergence of a gel, and to establish the gelation time, is to look at the moments of the cluster mass distribution. In the particular case of $\lambda=2$ the equations for (integer) moments are {\em recurrent}, and hence solvable. For instance,
\begin{equation}
\label{M234}
\begin{split}
&\frac{dM_2}{dt} = 1+2M_2^2 \\
&\frac{dM_3}{dt} = 1+6M_2 M_3 \\
&\frac{dM_4}{dt} = 1+2M_2^2+12M_2 M_4 
\end{split}
\end{equation}
Solving these equations one obtains
\begin{equation}
\label{M234:sol}
\begin{split}
&M_2 = \frac{\tan T}{\sqrt{2}} \\
&M_3 = \frac{\tan T+ \tfrac{2}{3} (\tan T)^3}{\sqrt{2}}\\
&M_4 = \frac{\frac{3T}{(\cos T)^6} + 5\tan T+ 8(\tan T)^3 + 3(\tan T)^5}{8\sqrt{2}}
\end{split}
\end{equation}
where we again used the shorthand notation $T=t\sqrt{2}$. All the moments diverge at $T_c=\pi/2$. This determines the gelation time: $T_c=\pi/2$, or $t_c=\pi/\sqrt{8}$. Using \eqref{M234:sol} we establish the divergence of the moments 
\begin{equation}
\label{M234:gel}
\begin{split}
&M_2 \simeq \frac{1}{2}\,(t_c-t)^{-1} \\
&M_3 \simeq \frac{1}{6}\,(t_c-t)^{-3} \\
&M_4 \simeq \frac{3\pi\sqrt{2}}{256}\,(t_c-t)^{-6} 
\end{split}
\end{equation}
near the gelation time. Generally 
\begin{equation}
\label{Mn:gel}
M_n\sim (t_c-t)^{-n(n-1)/2}
\end{equation}
for the $n^\text{th}$ moment. Therefore the moments of the cluster mass distribution exhibit a multi-scaling behavior. 

The divergence of the moments \eqref{Mn:gel} and multi-scaling also arise in the un-driven case \cite{EP03} and they can be understood using similar arguments. The idea is to use \eqref{GP:cmt} with $\lambda=2$. The auxiliary time variable is now 
\begin{equation}
\label{GP:tau-2}
\tau = \int_0^t dt'\,M_2(t')= -\frac{1}{2}\, \ln(\cos T)
\end{equation}
Since the auxiliary time diverges as $T\to T_c=\pi/2$, we can use continuum approaches; in particular, we can rely on the partial-differential equation \eqref{GP:cmt}. As explained in Ref.~\cite{EP03}, the cluster mass distribution attains (for large masses) a log-normal form
\begin{equation}
\label{log-normal} 
c_m(\tau)\sim e^{-\tau/4} m^{-3/2} \exp\!\left[-\frac{(\ln m)^2}{4\tau}\right]
\end{equation}
As a consistency check, let us compute the mass density. We write $\kappa = \ln m$ and find
\begin{equation*}
\sum_{m\geq 1}mc_m \sim e^{-\tau/4} \int_0^\infty d\kappa\,\exp\!\left[\frac{\kappa}{2}-\frac{\kappa^2}{4\tau}\right]
                                    \to    \text{const} 
\end{equation*}
as $\tau\to\infty$. The mass density is $M=t$, so it does approach a finite value at the gel point: $M\to t_c=\pi/\sqrt{8}$ 
Similarly, the $n^\text{th}$ moment grows as
\begin{eqnarray*}
\sum_{m\geq 1}m^n c_m &\sim&
 e^{-\tau/4} \int_0^\infty d\kappa\,\exp\!\left[\kappa\left(n-\frac{1}{2}\right)-\frac{\kappa^2}{4\tau}\right]\\
                                         &\sim& e^{n(n-1)\tau}
\end{eqnarray*}
This reduces to \eqref{Mn:gel} once we notice [see \eqref{GP:tau-2}] that 
\begin{equation}
\label{tau:t}
e^{2\tau}= \frac{1}{\cos T} \simeq \frac{1}{\sqrt{2}}\,(t_c-t)^{-1}
\end{equation}

The moments with $n\geq 1$, and even non-integer moments with $n>\frac{1}{2}$, behave according to \eqref{Mn:gel}. The zeroth moment, i.e., the total cluster density, exhibits a special behavior. This becomes evident once we note that $\sum_{m\geq 1}m^{-3/2}$ converges. Therefore $N\sim  e^{-\tau/4}$. Using \eqref{tau:t} we conclude that
\begin{equation}
\label{N:extreme-gel}
N(t) \sim (t_c-t)^{1/8}
\end{equation}
Vanishing of the total cluster density at $t_c$ shows that gelation is complete, namely the entire mass of the system is suddenly transformed into gel. 

\section{Exchange Processes Driven by a Localized Source}
\label{sec:local}

In previous sections we considered mass exchange processes supplemented by homogeneous input. Here we investigate the behaviors when input is spatially localized. The densities $c_m(\mathbf{r},t)$ then evolve according to a  system of partial differential equations 
\begin{eqnarray}
\label{loc:cmt}
\frac{\partial c_m}{\partial t} &=& \sum_{i,j} c_i c_j K_{i,j} \left[\delta_{m,i+1}+\delta_{m,i-1}-2\delta_{m,i}\right]
 \nonumber\\
&+&  D_m\nabla^2 c_m  + J \delta_{m,1}\delta(\mathbf{r})
\end{eqnarray}
The first term on the right-hand side of Eq.~\eqref{loc:cmt} accounts for exchange events. In writing the second term we have assumed that the transport mechanism is diffusion;  the diffusion coefficients $D_m$ generally depend on the mass $m$. The last term represents the localized source, namely the constant flux of monomers (with intensity $J$) at the origin. It suffices to model flux with delta function $\delta(\mathbf{r})$ since we are interested in distances greatly exceeding the size of the region where the monomers are injected. 

An infinite system of coupled non-linear partial differential equations \eqref{loc:cmt} possesses an outstanding challenge. Similar equations have been studied in the context of aggregation with a localized source \cite{Sid89,PLK:source,PLK:3particle}, and models with mass-independent diffusion coefficients proved to be much more amenable to analyses. For exchange processes with spatially varying densities, the situation is the same. One can already suspect this by noting that when $D_m=D= \text{const}$, the mass density satisfies a closed partial differential equation, the diffusion equation with a localized source:
\begin{equation}
\label{loc:M}
\frac{\partial M}{\partial t} =  D \nabla^2 M + J \delta(\mathbf{r})
\end{equation}
Hence the mass density is known and totally independent on the mass exchange:
\begin{equation}
\label{loc:M-sol}
M(\mathbf{r},t) = J\int_0^t \frac{dt'}{(4\pi D t')^{d/2}}\,\exp\!\left[-\frac{r^2}{4Dt'}\right]
\end{equation}
in $d$ dimensions. 

We limit ourselves to the physically most relevant three-dimensional case. In three dimensions, the mass density is asymptotically time-independent:
\begin{equation}
\label{loc:M-3d}
M = \frac{J}{4\pi D r}
\end{equation}

More precisely, the mass density is described by the Coulomb law \eqref{loc:M-3d} when $r\ll \sqrt{Dt}$. In the following we consider physically most important three-dimensional case and focus on the region $r\ll \sqrt{Dt}$ where not only the mass density but all cluster densities are stationary. 

We now analyze two basic reaction kernels, the product kernel and the constant kernel, and then mention chief behaviors for the generalized product kernels. At the end of this section we show that scaling approaches still apply to models in which diffusion coefficients vary with mass algebraically (specifically, we discuss models with $D_m\sim m^{-1}$). 

\subsection{Product Kernel}
\label{sec:loc_PK}

For the product kernel $K_{i,j}=Kij$, the governing equations in the stationary regime read
\begin{eqnarray}
\label{loc:cm}
0 &=& \frac{\sigma}{r}[(m+1)c_{m+1}-2mc_m+(m-1)c_{m-1}]  \nonumber\\
&+&  \nabla^2 c_m  + \frac{J}{D}\, \delta_{m,1}\delta(\mathbf{r})
\end{eqnarray}
where
\begin{equation}
\label{sigma}
\sigma = \frac{JK}{4\pi D^2}
\end{equation}
In this section, dimension-full quantities are used (apart from the mass of monomers which is still set to unity). The densities have dimension of the inverse volume, so $[c_m]=L^{-3}$ when $d=3$. The dimensions of the main parameters of the problem are
\begin{equation}
[J] = T^{-1},\quad [D] = L^2 T^{-1},\quad [K] = L^3 T^{-1}
\end{equation}
Thus $\sigma$ has dimension of the inverse length: $[\sigma]=L^{-1}$. 

For large masses we can treat $m$ as a continuous variable. The density $c=c_m(r)$ satisfies 
\begin{equation}
\label{loc:cmr}
\frac{\sigma}{r}\,\frac{\partial^2 (mc)}{\partial m^2}+ \nabla^2 c = 0
\end{equation}
Balancing terms gives $\frac{\sigma}{mr}\sim \frac{1}{r^2}$, so the typical mass increases linearly with distance: $m \sim \sigma r$. This together with the Coulomb law \eqref{loc:M-3d} for the mass density lead to the scaling form for the cluster mass distribution:
\begin{equation}
\label{loc:scaling}
c_m(r) = \frac{D\sigma^2}{K}\, (\sigma r)^{-3} F(\mu), \qquad \mu = \frac{m}{\sigma r}
\end{equation}
The scaling form \eqref{loc:scaling} is compatible with the Coulomb law \eqref{loc:M-3d} when 
\begin{equation}
\label{loc:FM-1}
\int_0^\infty d\mu\,\mu F(\mu)=1
\end{equation}

One can re-write \eqref{loc:scaling} in a manifestly dimensionless form
\begin{equation*}
\frac{c_m(r)}{\sigma^3} = \frac{D}{K\sigma}\, (\sigma r)^{-3} F(\mu)
\end{equation*}
In this and many other equations the distance $r$ from the origin naturally appears as a dimensionless combination $\sigma r$. The dimensionless parameter 
\begin{equation*}
\frac{D}{K\sigma} = 4\pi\,\frac{D^3}{JK^2}
\end{equation*}
provides the proper measure of the strength of the flux. 

By inserting \eqref{loc:scaling} into \eqref{loc:cmr} we obtain
\begin{equation}
\label{loc:FM}
(\mu^2+\mu)F''+(6\mu + 2)F+6F = 0
\end{equation}
The general solution of this equation is 
\begin{equation*}
F(\mu) = \frac{C_1+ C_2\left(\mu+2\ln\mu - \mu^{-1}\right)}{(\mu+1)^3}
\end{equation*}
This scaled mass distribution has the small mass singularity unless $C_2=0$. Equation \eqref{loc:FM-1} fixes another amplitude: $C_1=2$. Thus $F= 2(\mu+1)^{-3}$ and \eqref{loc:scaling} becomes
\begin{equation}
\label{loc:cm-sol}
c_m(r) = \frac{2D\sigma^2}{K}\, \frac{1}{(m+\sigma r)^3}
\end{equation}
In particular, the density of monomers is
\begin{equation}
\label{loc:mon-sol}
c_1(r) = \frac{2D}{K\sigma}\,r^{-3}
\end{equation}
The total cluster density is also given by a neat formula
\begin{equation}
\label{loc:N-sol}
N(r) = \frac{D}{K}\,r^{-2}
\end{equation}

The asymptotic behaviors \eqref{loc:mon-sol}--\eqref{loc:N-sol} are valid far away from the origin, more precisely when $r\gg \sigma^{-1}$. This is already evident from \eqref{loc:cm-sol}. Another way to see this is to consider the governing equation for $N(r)$ which is obtained by summing \eqref{loc:cm} for all $m\geq 1$:
\begin{equation}
\label{loc:N-eq}
0 =-\frac{\sigma}{r} c_1 + \nabla^2 N  + \frac{J}{D}\,\delta(\mathbf{r})
\end{equation}
One immediately verifies that \eqref{loc:mon-sol}--\eqref{loc:N-sol} agree with \eqref{loc:N-eq}. Near the origin, the source term dominates over reaction, so the total density is $N=J/(4\pi D r)$. Comparing this result with 
\eqref{loc:N-sol} we see that the crossover indeed occurs when $r\sim \sigma^{-1}$.  

The total cluster density is given by \eqref{loc:N-sol} up to distance of the order of $\sqrt{Dt}$ and it quickly vanishes for larger distances. Hence $\mathcal{N}(t)\sim \int_0^{\sqrt{Dt}} dr\,4\pi r^2 N(r)$ estimates 
the total number of clusters in the system. Using \eqref{loc:N-sol} we obtain
\begin{equation}
\label{loc:all}
\mathcal{N}\sim \frac{D}{K}\, \sqrt{Dt}
\end{equation}

\subsection{Constant Kernel}
\label{sec:loc_CK}

For the constant reaction kernel, $K_{i,j}=K$, the governing equations in the stationary regime become
\begin{equation}
\label{loc:const}
0 \!=\!  NK[c_{m+1}-2c_m+c_{m-1}]  +D\nabla^2 c_m  + J \delta_{m,1}\delta(\mathbf{r})
\end{equation}
In the continuum limit the density $c=c_m(r)$ satisfies 
\begin{equation}
\label{loc:cmr-const}
\frac{NK}{D}\,\frac{\partial^2 c}{\partial m^2}+ \nabla^2 c = 0
\end{equation}
Balancing the terms of this equation yields $\frac{NK}{D m^2}\sim \frac{1}{r^2}$, leading to $m\sim\sqrt{NKr^2/D}$. Another estimate of the typical mass is $m\sim M/N$. Balancing these estimates and using \eqref{loc:M-3d} we get 
\begin{equation}
\label{loc:N-estimate}
N \sim \frac{D}{K}\,\sigma^{2/3} r^{-4/3}, \quad m\sim \sigma^{1/3} r^{1/3}
\end{equation}

These estimates determine an appropriate scaling form for the cluster mass distribution, e.g., as a scaled mass one should use $m/(\sigma r)^{1/3}$. Putting such a form into \eqref{loc:cmr-const} and solving the resulting ordinary differential equation one finds the scaled mass distribution. The results are less cumbersome if we write $c_m(r)$ as 
\begin{equation}
\label{loc:scaling-const}
c_m(r) = \frac{D\alpha^2}{K}\,(\alpha r)^{-5/3}\, F(\mu), \qquad \mu = \frac{m}{(\alpha r)^{1/3}}
\end{equation}
where instead of $\sigma$ defined by Eq.~\eqref{sigma} we use 
\begin{equation}
\alpha = \frac{36\sigma}{\pi} = \frac{9JK}{\pi^2 D^2}
\end{equation}
The scaled mass distribution then reads 
\begin{equation}
\label{loc:FM-const}
F(\mu) = \frac{4}{9}\,\frac{\mu}{(\mu^2+1)^3}
\end{equation}
The density of monomers is
\begin{equation}
\label{loc:mon-const}
c_1(r) = \frac{4D}{9K}\,r^{-2}
\end{equation}
and the total cluster density is 
\begin{equation}
\label{loc:N-const}
N(r) = \frac{D}{9K}\,\frac{\alpha^{2/3}}{r^{4/3}}
\end{equation}
Near the origin the total density is $N=J/(4\pi D r)$ which again crosses over to \eqref{loc:N-const} at $r\sim \sigma^{-1}$.  

Using \eqref{loc:N-const} we find an estimate for the total number of clusters in the system:
\begin{equation}
\label{loc:all-const}
\mathcal{N}\sim \frac{D}{K}\, \sigma^{2/3} (Dt)^{5/6}
\end{equation}

\subsection{Generalized Product Kernels}

In the continuum limit the density $c=c_m(r)$ satisfies 
\begin{equation}
\label{loc:cmr-GPK}
\frac{M_\lambda K}{D}\,\frac{\partial^2 (m^\lambda c)}{\partial m^2}+ \nabla^2 c = 0
\end{equation}
The same arguments as in the previous subsections lead to the scaling form
\begin{equation}
\label{loc:scaling-GPK}
c_m(r) = \frac{D}{K}\,\frac{\sigma^2}{(\sigma r)^{1+2\Lambda}}\, F(\mu), \qquad \mu = \frac{m}{(\sigma r)^\Lambda}
\end{equation}
where $\Lambda=1/(3-2\lambda)$. We omit a rather complicated solution for the scaled mass distribution and only mention the asymptotic for the total cluster density
\begin{equation}
\label{loc:N-GPK}
N(r) \sim \frac{D}{K}\,\frac{\sigma^{1-\Lambda}}{r^{1+\Lambda}}
\end{equation}
and for the total number of clusters in the system
\begin{equation}
\label{loc:all-GPK}
\mathcal{N}\sim \frac{D}{K \sigma}\,(\sigma^2 Dt)^{1-\Lambda/2}
\end{equation}

Equations \eqref{loc:scaling-GPK}--\eqref{loc:all-GPK} provide a proper description only if the total number of clusters diverges in the $t\to\infty$ limit. This together with \eqref{loc:all-GPK} give $\Lambda<2$ implying that the index $\lambda$ of the generalized product kernels must obey $\lambda<\frac{5}{4}$. Indeed, the rate equation framework which was used in the derivations is deterministic, so its applicability is based on the assumption that the total number of clusters in the system is infinite. For systems with finite number of clusters, there are always fluctuations. In the exchange processes with a localized source the total number of clusters is of course finite, but since we are interested in the large time limit the emergent results are consistent only when $\lim_{t\to\infty} \mathcal{N}(t)=\infty$. 

\subsection{Models with $D_m=m^{-1}$}
\label{sec:inverse}

The assumption that the diffusion coefficients do not depend on mass was crucial in the preceding analysis. The diffusion coefficients generally decrease with mass, however. The decay is usually algebraic, $D_m\sim m^{-\nu}$; see \cite{einstein} for arguments in favor of specific values of the mobility exponent (e.g. $\nu=1$ and $\nu=3/2$ in some problems involving two-dimensional clusters). It turns out the models with mass-dependent algebraically decaying diffusion coefficients are still tractable. The idea is to assume the emergence of a stationary mass distribution and the validity of scaling. Whenever the results are consistent, one hopes that they are asymptotically exact.   

For concreteness, let us look at a set of models with diffusion coefficients inversely proportional to the mass, $D_m=D/m$. For such models, the mass density varies as
\begin{equation}
\label{loc:MN-inverse}
\frac{\partial M}{\partial t} =  D \nabla^2 N + J \delta(\mathbf{r})
\end{equation}
In the long time limit the left-hand side of Eq.~\eqref{loc:MN-inverse} tends to zero, so the cluster density is 
\begin{equation}
\label{loc:Coulomb}
N = \frac{J}{4\pi D r}
\end{equation}
(Recall, that we consider the three-dimensional case.) We know $N(r)$, so the model with constant kernel is the simplest. In that model, the densities satisfy
\begin{equation}
\label{loc:inverse-steady}
0 = \frac{\sigma}{r}[c_{m+1}-2c_m+c_{m-1}]  + \nabla^2 \frac{c_m}{m}  + \frac{J}{D}\, \delta_{m,1}\delta(\mathbf{r})
\end{equation}
The scaling solution is found using the same techniques as previously to yield
\begin{equation}
\label{loc:cm-inverse}
c_m(r) = \frac{2D\sigma^2}{K}\, \frac{m}{(m+\sigma r)^3}
\end{equation}

The monomer density is given by exactly the same formula \eqref{loc:mon-sol} as in the model with product kernel and mass-independent diffusion coefficients, but other densities differ. Using \eqref{loc:cm-inverse}, we compute the mass density 
\begin{equation}
\label{loc:M-inverse}
M(\mathbf{r},t) = \frac{2D\sigma^2}{K} \left[\int_{\sigma r}^{m_\text{max}} \frac{dz}{z}-\frac{3}{2}\right]
\end{equation}
An estimate $r\sim\sqrt{Dt/m_\text{max}}$ gives $m_\text{max}\sim Dt/r^2$. Thus
\begin{equation}
\label{loc:M-log}
M(r,t) = \frac{2D\sigma^2}{K} \left[\ln(\sigma^2 Dt)- 3\ln(\sigma r)-\frac{3}{2}\right]
\end{equation}

Models with $D_m=D/m$ and non-constant kernels are more complicated. Let's try, however, to probe chief behaviors relying on scaling and assuming the emergence of a stationary mass distribution. For models with generalized product kernel we arrive at a familiar scaling form
\begin{equation}
\label{loc:inverse-gen}
c_m(r) = \frac{D}{K}\,\frac{\sigma^2}{(\sigma r)^{1+\delta}} F(\mu), \qquad \mu = \frac{m}{(\sigma r)^\delta}
\end{equation}
with $\delta = 1/(1-2\lambda)$. Even the the mass distribution is stationary, the mass density may be non-stationary as it is already exemplified by Eq.~\eqref{loc:M-inverse} in the case of $\lambda=0$. Using the scaling form \eqref{loc:inverse-gen} we arrive at an estimate of the mass density
\begin{equation}
\label{loc:M-steady}
M(r) \sim \frac{D}{K} \sigma^{\frac{2-2\lambda}{1-2\lambda}}r^{\frac{2\lambda}{1-2\lambda}}
\end{equation}
The dependence of $M(r)$ on the distance suggests that the mass density is stationary for $\lambda<0$. Using mass conservation, $\int_0^{r_\text{max}}dr\,r^2 M(r)\sim Jt$, we conclude that \eqref{loc:M-steady} is valid for $r<r_\text{max}$ with
\begin{equation}
\label{loc:max-1}
\sigma r_\text{max}\sim (\sigma^2 Dt)^{\frac{1-2\lambda}{3-2\lambda}}
\end{equation}
and the mass density quickly vanishes for $r>r_\text{max}$. 

A more rigorous derivation of \eqref{loc:M-steady} in the $\lambda<0$ region could be obtained if we determine the scaled mass distribution. By inserting the scaling form \eqref{loc:inverse-gen} into
\begin{equation}
\frac{M_\lambda K}{D}\,\frac{\partial^2 (m^\lambda c)}{\partial m^2}+ \nabla^2 \frac{c}{m} = 0
\end{equation}
we obtain
\begin{equation}
\label{loc:FM-inv}
\frac{1}{2}\,\mu F'' + (2-\lambda) F' + \frac{1-\lambda}{\mu}\,F + I_\lambda (\mu^\lambda F)''=0 
\end{equation}
with $I_\lambda=\frac{1}{2}(1-2\lambda)^2\int_0^\infty d\mu\,\mu^\lambda F(\mu)$.  Apart from the already solved case of $\lambda=0$, Eq.~\eqref{loc:FM-inv} apparently does not have solutions in terms of known special functions. Still, the large mass limit can be probed analytically. In this limit, the last term in Eq.~\eqref{loc:FM-inv} is negligible. 
Dropping this term and solving the resulting equation we obtain
\begin{equation}
\label{loc:FM-inv-sol}
F \sim \mu^{-2(1-\lambda)}\quad\text{for}\quad \mu\gg 1
\end{equation}
For $\lambda<0$, the integral $\int^\infty d\mu\,\mu F(\mu)$ converges and hence the mass density does scale according to \eqref{loc:M-steady}. 

When $0<\lambda< \frac{1}{2}$, the mass density is non-stationary:
\begin{eqnarray}
\label{loc:M-steady-2}
M (r,t) &\sim&  \frac{D\sigma^2}{K}\,(\sigma r)^{\delta -1} \int_0^{\mu_\text{max}} d\mu\,\mu F(\mu)\nonumber\\
&\sim&  \frac{D\sigma^2}{K}\,(\sigma r)^{\delta -1} \int_0^{Dt/[r^2(\sigma r)^\delta]} 
d\mu\,\mu^{2\lambda-1}\nonumber\\
&\sim&  \frac{D\sigma^2}{K}\left(\!\frac{Dt}{r^2}\right)^{2\lambda}
\end{eqnarray}
where we have used an estimate $m_\text{max}\sim Dt/r^2$. Also for $0<\lambda< \frac{1}{2}$, Eq.~\eqref{loc:M-steady-2} is valid for $r<r_\text{max}$ with
\begin{equation}
\label{loc:max-2}
\sigma r_\text{max}\sim (\sigma^2 Dt)^{(1-2\lambda)/(3-4\lambda)}
\end{equation}

Using \eqref{loc:Coulomb} together with the estimates \eqref{loc:max-1} and \eqref{loc:max-2} of the maximal distance we find that the total number of clusters scales as
\begin{equation}
\label{loc:all-inv}
\mathcal{N}\sim \frac{D}{K \sigma}\,(\sigma^2 Dt)^{\beta}
\end{equation}
with
\begin{equation*}
\beta = 2(1-2\lambda)\times
\begin{cases}
(3-2\lambda)^{-1}  & \lambda<0\\
(3-4\lambda)^{-1}  & 0<\lambda<\frac{1}{2}
\end{cases}
\end{equation*}

The above predictions are inconsistent when $\lambda\geq \frac{1}{2}$. The most natural guess is that the cluster mass distribution does not reach a stationary regime in this region. It would be interesting to develop a quantitative understanding of the behavior in the $\lambda\geq \frac{1}{2}$ region.

\section{Conclusions and Discussion}
\label{sec:conl}

We studied mass exchange processes driven by a source of monomers. We considered both homogeneous and localized input. For processes with homogeneous source, we employed scaling methods and analyzed in detail three basic exchange kernels---the product kernel, the sum kernel, and the constant kernel; we also probed a one-parameter family of generalized product kernels. We also exactly solved the mass exchange process with product reaction kernel supplemented by the homogeneous source of monomers. In all these systems, a steady state is never reached. A generalization of our approach to exchange processes with kernels $K_{i,j}=i^aj^b+i^bj^a$ and homogeneous input seemingly leads to the same conclusion. It would be interesting to find a general proof that steady states are impossible in exchange processes driven by a homogeneous source. 

We showed that the mass exchange processes with spatially localized input can be {\em simpler} than the corresponding processes with homogeneous input. This is obvious e.g. by comparing the cluster mass distributions for the exchange process with product kernel: Eqs.~\eqref{cmt:scaling} and \eqref{F:sol_B} for the homogeneous input and Eq.~\eqref{loc:cm-sol} for the localized input. The chief reason is that the cluster mass distribution is often stationary when the source is spatially localized. We demonstrated that stationary mass distributions naturally emerge for exchange processes with mass-independent diffusion coefficients, e.g., for exchange processes with generalized product kernel, $K_{i,j}=(i j)^\lambda$ with $\lambda<\frac{5}{4}$, where the bound on the homogeneity index assures that the deterministic rate equation description is asymptotically relevant. 

We outlined possible behaviors for the mass exchange processes with spatially localized input in the situations when diffusion coefficients depend on mass. Specifically, we analyzed models in which diffusion coefficients are inversely proportional to the mass. We presented evidence that the cluster size distribution may not reach a stationary state. This depends on the details of the exchange kernel; e.g., for exchange processes with generalized product kernel a stationary state is apparently never reached when $\lambda>\frac{1}{2}$. The behaviors in the continuously evolving regime are the challenge for future work. 

The power of scaling in the context of exchange processes is impressive. Another general lesson is that seemingly intractable systems of infinitely many coupled non-linear partial differential equations for spatially varying densities can be simpler than their homogeneous counter-parts. This is not clear a priori, and even the emergence of the stationary distribution does not make the spatial problem manifestly simpler as it involves the Laplace operator rather than the first derivative in time. 

Finally we mention that exchange processes often occur in low spatial dimensions where the rate equation approach is invalid. It would be interesting to analyze simplest exchange processes in one dimension.


\begin{thebibliography}{99}
\bibitem{similarity}
     This similarity appears even on a formal level, e.g., the exact solution \eqref{prod:exact} is also the exact solution 
     for the constant-kernel aggregation process.
\end{thebibliography}
\end{document}